\documentstyle[12pt]{article}
\newlength{\aivwidth}   
\newlength{\tmpwidth}   

\def\beq{\begin{equation}}
\def\eeq{\end{equation}}
\def\beqar{\begin{eqnarray}}
\def\eeqar{\end{eqnarray}}
\def\barr#1{\begin{array}{#1}}
\def\earr{\end{array}}
\def\bfi{\begin{figure}}
\def\efi{\end{figure}}
\def\btab{\begin{table}}
\def\etab{\end{table}}
\def\bce{\begin{center}}
\def\ece{\end{center}}
\def\nn{\nonumber}

\def\text{\textstyle}

\def\De{\Delta}

\def\refeq#1{\mbox{(\ref{#1})}}

\def\citere#1{\mbox{Ref.~\cite{#1}}}
\def\citeres#1{\mbox{Refs.~\cite{#1}}}

\renewcommand{\O}{{\cal O}}

\def\mathswitchr#1{\relax\ifmmode{\mathrm{#1}}\else$\mathrm{#1}$\fi}

\newcommand{\PW}{\mathswitchr W}
\newcommand{\PZ}{\mathswitchr Z}

\newcommand{\PH}{\mathswitchr H}

\def\mathswitch#1{\relax\ifmmode#1\else$#1$\fi}

\newcommand{\MW}{\mathswitch {M_\PW}}

\newcommand{\MZ}{\mathswitch {M_\PZ}}
\newcommand{\MH}{\mathswitch {M_\PH}}

\newcommand{\phrd}[1]{Phys.\ Rev.\ {\bf D#1}}

\newcommand{\phrl}[1]{Phys.\ Rev.\ Lett.\ {\bf #1}}
\newcommand{\nphb}[1]{Nucl.\ Phys.\ {\bf B#1}}

\newcommand{\phlb}[1]{Phys.\ Lett.\ {\bf B#1}}

\newcommand{\aph}[1]{Ann.\ Phys.\ (NY) {\bf #1}}

\def\tr#1{\,\mbox{tr}\left\{#1\right\}}

\newcommand{\spc}{\phantom{{}+{}}}

\newcommand{\lag}{{\cal {L}}}
\newcommand{\leff}{\lag_{\rm eff}}

\newcommand{\hH}{\hat{H}}
\newcommand{\hW}{\hat{W}}

\newcommand{\hB}{\hat{B}}

\newcommand{\hV}{\hat{V}}
\renewcommand{\ss}{\scriptscriptstyle}

\newcommand{\OH}[1]{\O(\MH^{#1})}
\newcommand{\dpp}{(x,\partial_x+i p)}

\newcommand{\nl}{\nn\\&&{}}
\newcommand{\tde}{\De_{\MH}}
\newcommand{\fac}{\frac{1}{16\pi^2}}

\begin{document}
\title{One-Loop Effects\\ of a Heavy Higgs Boson:\\
a Functional Approach\thanks{To appear in the proceedings of the
International Symposium on Vector Boson Self-Interactions,
Feb.~1--3, UCLA}}
\author{Stefan Dittmaier$^1$ and Carsten Grosse-Knetter$^2$\\
\normalsize $^1$University of Bielefeld, Bielefeld, Germany\\
\normalsize $^2$University of Montreal, Montreal, Canada}
\date{BI-TP 95/14\\UdeM-GPP-TH-95-23\\hep-ph/9503440\\March 1995}

\begin{titlepage}
\maketitle

\begin{abstract}
We integrate out the Higgs boson in the electroweak standard model at
one loop, assuming that it is very heavy. We construct a low-energy
effective Lagrangian, which parametrizes the one-loop effects of the
heavy Higgs boson at $\OH{0}$. Instead of applying conventional
diagrammatical techniques, we integrate out the Higgs boson directly
in the path integral.
\end{abstract}
\end{titlepage}

\section*{Introduction}
Effective Lagrangians are used in order to describe the low-energy
effects of heavy particles. An effective Lagrangian
contains only light particles, and the heavy particles' effects
are parametrized in terms of effective interactions of the light
ones.

An effective Lagrangian can be constructed from the underlying
theory by integrating out the heavy particles. This can be done in
two different ways:
\begin{itemize}
\item {\em The diagrammatical method:\/} One calculates the relevant
Feynman graphs
with heavy particles and matches the full theory to the effective
one.
\item {\em The functional method:\/} One integrates out the the heavy fields
directly in the path integral.
\end{itemize}
Here we focus on the functional method which turns out to be much
more elegant and simpler than the diagrammatical one.

As a phenomenologically important application we consider the
electroweak standard model (SM) provided that $\MH \gg\MW, E$. We
integrate out the Higgs boson and determine the formal limit
$\MH\to\infty$ of the SM, i.e.\  all contributions of the
Higgs boson to the resulting
effective Lagrangian at $\OH{0}$ (which includes
$\log\MH$-terms).

In these proceedings we can sketch our method and our results only
very briefly. The reader who is interested in a
more detailed presentation
is referred to the original articles \citeres{sdcgk1,sdcgk2}.

\section*{Integrating out the Higgs field}
We briefly describe the basic concepts of our method to
integrate out heavy fields in the path integral.

\subsection*{The background-field method}
The SM Lagrangian contains terms cubic and quartic in the Higgs field.
Thus, the integral over the Higgs field is not of Gaussian type.
However, this problem can be circumvented by applying the
background-field method (BFM) \cite{bfm1,bfm2},
where each field is split into a
(classical) background field and a quantum field, such that the
functional integral is only performed over the latter. The background
fields correspond at the diagrammatical level to tree lines while the
quantum fields correspond to lines in loops.
Thus, at one loop it is sufficient to consider only the part of the
Lagrangian which is quadratic in the quantum fields. Then the heavy
quantum field can be integrated out by Gaussian integration.

\subsection*{The Stueckelberg formalism}
Another advantage of the use of the BFM is that different gauges
may be chosen for the quantum and  background fields,
respectively \cite{bfm1,bfm2}. For our purposes it is useful
to choose a generalized $R_\xi$-gauge \cite{bfm2}
for the quantum fields (such
that the quantized Lagrangian is still invariant under gauge
transformations of the background fields \cite{bfm1,bfm2}) but the unitary
gauge for the background fields.
This aim can be achieved by applying a generalized \cite{sdcgk1,sdcgk2,chey}
Stueckelberg transformation \cite{stue1,stue2}
to the (background and
quantum) vector fields, which removes the background Goldstone fields
from the Lagrangian. After all calculations are  done, this
transformation is inverted in order to reintroduce the
background Goldstone
fields and to obtain a
manifestly gauge-invariant result.

\subsection*{Diagonalization of the Lagrangian}
The one-loop Lagrangian of the SM (i.e.\ the part quadratic in the
quantum fields)
contains terms linear in the quantum Higgs field $H$. After Gaussian
integration these would yield terms with inverse operators acting on
the quantum fields. However, one can apply appropriate shifts to the
quantum fields, such that these linear terms are removed while the
Higgs-independent part of the Lagrangian remains unaffected
\cite{sdcgk1,chey,gale}. The resulting Lagrangian can then be written
in the symbolic form
\beq
\lag^{\rm{1-loop}}=-\frac{1}{2}H\tilde{\Delta}_{\ss H}(x,\partial_x) H
+\left.\lag^{\rm{1-loop}}\right|_{H= 0},
\eeq
where the operator $\tilde{\Delta}_{\ss H}$, which depends on the
background fields, also contains
contributions from the terms originally linear in the quantum Higgs field
$H$ owing to the shifts.

\subsection*{{\boldmath${1}/{\MH}$}-expansion}
Next the Gaussian integration over the quantum Higgs field can be
performed. The resulting functional determinant can be parametrized in
terms of an effective Lagrangian \cite{sdcgk1,chan}
\beq
\leff=\frac{i}{2}\int\frac{d^4p}{(2\pi)^4}\log\left(\tilde{\Delta}_{\ss
H}\dpp\right).
\eeq
Then one can perform a Taylor expansion of the expression
$\tilde{\Delta}_{\ss H}\dpp$ around $\tilde{\Delta}_{\ss H}(x,ip)$
(derivative expansion) followed by a Taylor expansion of the
logarithm. These expansions yield one-loop vacuum
integrals of the type
\beq
\int d^4 p\frac{p_{\mu_1}\ldots p_{\mu_{2k}}}{(p^2-\MH^2)^l
(p^2-\xi M_{i}^2)^m},\qquad M_i=\MW,\MZ.
\eeq
These are $\OH{0}$ or higher only for $4+2(k-l-m)\ge 0$.
Thus, in the two above-mentioned Taylor expansions only a finite
number of terms contribute to
the effective
Lagrangian $\leff$ at $\OH{0}$.

\subsection*{Elimination of the background Higgs field}
After the integration over the quantum Higgs field $H$, the effective
Lagrangian still contains the background Higgs field $\hH$. The latter
corresponds
to Higgs tree lines, and thus can easily be eliminated by a
propagator expansion
Diagrammatically this means that the $\hH$
propagator shrinks to a point rendering such (sub-)graphs
irreducible, which contain background Higgs lines only. Equivalently, the
background Higgs-field can be eliminated by applying the classical
equations of motion which are valid at tree level.
Before
eliminating the field $\hH$, the Higgs sector has to be renormalized by
adding the Higgs-dependent part of the counterterm Lagrangian.

\section*{The heavy-Higgs limit of the standard model}
Proceeding as explained above, we find the formal limit of the
SM for \mbox{$\MH\to\infty$} -- i.e.\ the Lagrangian which contains
the non-decoupling ($\OH{0}$) effects of
the Higgs boson -- at one loop \cite{sdcgk2}:
\beq
\left.\lag^{\rm{1-loop}}_{\rm{SM}}\right|_{\MH\to\infty} =
\lag^{\rm{1-loop}}_{\rm{GNLSM}} +\leff.
\label{l1}
\eeq
In eq.~\refeq{l1} $\lag^{1-loop}_{GNLSM}$ is the one-loop Lagrangian
of the gauged nonlinear $\sigma$-model (GNLSM)
\cite{long}, which is obtained from the SM Lagrangian by simply
omitting the Higgs field in the unitary-gauge.
$\leff$ is the effective Lagrangian generated by
integrating out the Higgs field and parametrizes the one-loop effects
of the heavy Higgs field. Omitting
terms which do not contribute to the S-matrix we find \cite{sdcgk2}
\beqar
\leff^{\rm{S-matrix}}&=&
\spc\fac \frac{3}{8}\bigg(\tde+\frac{5}{6}\bigg)\frac{g_1^2}{g_2^2}
{\MW^2}\left(\tr{T \hV_\mu}\right)^2\nl
-\fac\frac{1}{24}\bigg(\tde+\frac{5}{6}\bigg)g_1g_2
\hB_{\mu\nu}\tr{T\hW^{\mu\nu}}\nl
+\fac\frac{1}{48}
\bigg(\tde+\frac{17}{6}\bigg)
i{g_1}\hB_{\mu\nu}\tr{T[\hV^\mu,\hV^\nu]}\nl
-\fac\frac{1}{24}\bigg(\tde+\frac{17}{6}\bigg)ig_2
\tr{\hW_{\mu\nu}[\hV^\mu,\hV^\nu]}\nl
-\fac\frac{1}{12}\bigg(\tde+\frac{17}{6}\bigg)
\left(\tr{\hV_\mu\hV_\nu}\right)^2\nl
-\fac\frac{1}{24}\bigg(\tde+\frac{79}{3}-\frac{27\pi}{2\sqrt{3}}\bigg)
\left(\tr{\hV_\mu\hV^\mu}\right)^2
+\OH{-2}
\label{l2}
\eeqar
with
\beq
\tde=\De-\log\left(\frac{\MH^2}{\mu^2}\right), \qquad
\De =\frac{2}{4-D}-\gamma_E+\log(4\pi),
\label{tde}
\eeq
where $D$ is the space-time dimension in dimensional regularization,
$\gamma_E$ is Euler's constant, and $\mu$ is the reference mass.

In eq.~\refeq{l2} we have used the notation of
\citeres{sdcgk2,long,hemo}, which we specify here only for
the case of the $U$-gauge, where the background Goldstone
fields are absent, and thus the physical content of the terms in
$\leff$ is most obvious:
\beq
\hV^\mu=-\frac{i}{2}\left(g_2\hW^\mu_i\tau_i+{g_1}
\hB^\mu \tau_3\right),\qquad T=\tau_3, \qquad
\hW^{\mu\nu}=\frac{1}{2}\hW^{\mu\nu}_i\tau_i,
\eeq
where $\hB^\mu$ and $\hW^\mu$  are the $U(1)$ and $SU(2)$ gauge
fields, respectivly, $g_1$ and $g_2$ are the corresponding gauge
couplings and the $\tau_i$ are the Pauli matrices. The hats over the
fields indicate that these are background fields.

The result of our functional calculation agrees with the result of
the diagrammatical calculation in \citere{hemo}.

The first two terms in eq.~\refeq{l2} contain vector-boson two-point
(and higher) functions, the next two three-point (and higher) functions
and the last two four-point functions. Thus, the first two
parametrize the effects of the heavy Higgs boson to LEP~1 physics,
the next two the effects to LEP~2 physics and the last two those to
LHC physics.

$\leff$ (\ref{l2}) does not contain custodial $SU(2)$
violating terms of dimension 4, although there are 7 such terms which
are by naive power counting
expected to be generated when integrating
out the Higgs field \cite{long,hemo}.
As shown in \citere{sdcgk2}, within our functional calculation it is
obvious that these terms vanish (while in a diagrammatical
calculation \cite{hemo} their absence seems to be an accidental
cancellation).

The effective interaction terms in eq.~\refeq{l2} have logarithmically
divergent coefficients. Owing to the renormalizability of the SM,
these UV-divergences cancel against the logarithmically divergent
one-loop contributions from the GNLSM Lagrangian in eq.~\refeq{l1}
\cite{long}. Since logarithmic divergences and $\log\MH$-terms in
$\leff$ always occur in the combination $\tde$
(\ref{tde}), the $\log\MH$-terms in the SM coincide with the
logarithmically divergent terms in the GNLSM,
as assumed in \citere{long}.
However, in addition eq.~\refeq{l2} contains finite and constant
differences between the SM and the GNLSM at one loop.

\section*{Conclusion}
Our purpose with this project is twofold: On the one hand, we
integrate out the Higgs boson in the electroweak standard model
at one loop. We
parametrize the non-decoupling (i.e. $\OH{0}$)
effects of this field in terms
of a low-energy effective Lagrangian.
On the other hand, we have
developed a functional method to integrate out heavy fields directly in
the path integral. This method can be applied to integrate out
any kind of
non-decoupling heavy field and also be generalized to a decoupling
scenario.

Our method is an alternative to the conventional
diagrammatical techniques and turns out to be a huge simplification.
While in a diagrammatical calculation various Green functions (i.e.\
very many Feynman diagrams) have to be calculated
and the effective Lagrangian has to be determined
indirectly by matching the
full theory to the effective one \cite{hemo}, in a
functional calculation the effective Lagrangian
is generated {\em directly\/}
by integrating out the heavy fields.
The use of the BFM and of the Stueckelberg formalism
automatically ensures gauge invariance of the result.

\end{document}